%% file: main.tex
\documentclass[fleqn,10pt]{wlscirep}
\usepackage[utf8]{inputenc}
\usepackage[T1]{fontenc}
\usepackage{graphicx} 
\usepackage{subcaption}
\usepackage{mathtools}
\usepackage{xcolor} 
\usepackage{comment} 
\usepackage{algorithm}
\usepackage[noend]{algpseudocode}
\usepackage{amsmath}
\usepackage{amsfonts} 
\usepackage{array}

\DeclareMathOperator*{\argmin}{\arg\!\min}

\algnewcommand{\LineComment}[1]{\State \(\triangleright\) #1}

\title{ Parallel Algorithms for Median Consensus Clustering in Complex Networks }

\author[1,*]{Md Taufique Hussain}
\author[2]{Mahantesh Halappanavar}
\author[2]{Samrat Chatterjee}
\author[3]{Filippo Radicchi}
\author[3]{Santo Fortunato}
\author[1,*]{Ariful Azad}

\affil[1]{Department of Intelligent Systems Engineering, Indiana University, Bloomington, IN USA}
\affil[2]{Data Sciences and Machine Intelligence Group, Pacific Northwest National Laboratory, Richland, WA USA}
\affil[3]{Center for Complex Networks and Systems Research (CNetS), Indiana University, Bloomington, IN USA}

\affil[*]{Corresponding authors: mth@iu.edu, azad@iu.edu}

\date{}

\begin{abstract}
    We develop an algorithm that finds the consensus of many different clustering solutions of a graph. 
    We formulate the problem as a median set partitioning problem and propose a greedy optimization technique. 
    Unlike other approaches that find median set partitions, our algorithm takes graph structure into account and finds a comparable quality solution much faster than the other approaches. 
    For graphs with known communities, our consensus partition captures the actual community structure more accurately than alternative approaches. 
    To make it applicable to large graphs, we remove sequential dependencies from our algorithm and design a parallel algorithm. 
    Our parallel algorithm achieves 35x speedup when utilizing 64 processing cores for large real-world graphs from single-cell experiments.
\end{abstract}

\begin{document}
\flushbottom
\maketitle
\thispagestyle{empty}

\input{intro_new}

\input{method}

\input{experiments}

\input{conclusion}

\bibliography{bibfile}

\input{Appendix}

\end{document}

%% file: intro_new.tex
\section{Introduction}
Complex networks often display community structure, where nodes are densely connected within communities but sparsely connected to nodes in other communities~\cite{fortunato2010community, porter2009communities}. 
Identifying communities (also called clusters or modules) in a network or graph has numerous applications in biology, social science, and various other scientific fields. 
Over decades of research, many community detection algorithms have been developed, each optimizing different objective functions~\cite{coscia2011classification, newman2004finding}. 
Due to the absence of a universally accepted definition of communities, different algorithms often identify significantly different clusters on the same graph~\cite{dao2020community, fortunato2016community}. 
Moreover, many algorithms are stochastic, leading to variations in clusters depending on input parameters and approximation techniques. 
These variabilities make it challenging to select one clustering over others.

Instead of choosing a single partition from multiple input partitions, we can combine many input partitions to create a new one. This consensus approach is a valuable way to cluster data, especially when no single metric can determine the best partition. 
Consensus clustering is a well-established technique in data analysis aimed at finding the median (or consensus) partition~\cite{vega2011survey, nguyen2007consensus, goder2008consensus}. 
The median partition is the one most similar, on average, to all input partitions, based on between-partition similarity measures such as normalized mutual information (NMI), Rand index, adjusted Rand index, Fawlkes and Mallows measure, variation of information, split-join measure, and the Jaccard index.
In its standard form, this combinatorial optimization problem is NP-complete~\cite{filkov2004integrating}, making it infeasible to solve for moderate to large datasets. Consequently, existing methods employ various greedy optimizations.

Despite being well-studied in data science, consensus clustering has received less attention in network science. 
Most existing graph consensus clustering methods overlook the edges in the graph and treat the problem as a set partition problem. 
For instance, Lancichinetti and Fortunato use a consensus matrix based on the co-occurrence of vertices in clusters from the input partitions, repeatedly applying the original graph clustering algorithm until the consensus matrix stabilized~\cite{lancichinetti2012consensus}.
This method is impractical for large graphs due to the extensive memory and runtime required to handle all possible pairs of vertices.
Tandon et al. propose an improvement by considering small subsets of vertex pairs instead of all possible pairs~\cite{tandon2019fast}. 
However, both approaches struggle to converge to a solution when the input clusterings exhibit high variability, such as those produced by different classes of algorithms. Recent works point out that obtaining a single consensus partition may not be appropriate, if there are multiple high-quality solutions with significantly different clusters~\cite{peixoto2021revealing,kirkley2022representative, jeub2018multiresolution}. 

In this paper, we develop a novel graph consensus clustering algorithm that minimizes the total distance between the consensus partition and all input partitions. 
While there are various ways to define inter-partition distance~\cite{meila2005comparing}, the Mirkin distance is by far the most used in the context of consensus clustering~\cite{filkov2004integrating} and we use it here as well. 
For this optimization, we introduce a greedy algorithm that leverages the graph structure for iterative improvement of the consensus.
We empirically demonstrate that our structure-aware algorithm produces a consensus partition that reduces the average distance between the consensus and all input partitions. 
For graphs with known communities, our consensus partition captures the actual community structure more accurately than alternative approaches. 
Additionally, unlike the graph-agnostic solution introduced by Filkov and Skiena~\cite{filkov2004integrating}, our algorithm avoids quadratic memory requirements, making it suitable for large graphs. 
We eliminate sequential dependencies between successive consensus updates and develop a parallel algorithm for multi-core processors. 
Our parallel algorithm runs 35 times faster than the sequential algorithm when using 64 processing cores.

%% file: method.tex
\section{Methods}

\subsection{The consensus clustering problem}
\label{subsec:problem}
Let $G(V, E)$ be a graph with a set of $n$ vertices $V$ and a set of $m$ edges $E$.
A partition or clustering of $G$ is a collection of non-overlapping clusters, wherein each vertex is exclusively assigned to a single cluster.
Let $\textbf{P}_1, \textbf{P}_2, ..., \textbf{P}_k$ denote $k$ partitions of the graph.
Consequently, a vertex $v$ is included in $k$ clusters, one cluster from each partition.
Let $d(\textbf{P}_i, \textbf{P}_j)$ represent the distance between the partitions $\textbf{P}_i$ and $\textbf{P}_j$. 
Suppose $\textbf{C}$ is a consensus of the $k$ input partitions. The optimal consensus partition, $\textbf{C}^*$, is the one that minimizes the
sum of the distances between the consensus and all input partitions:
\begin{equation}
     \textbf{C}^* = \argmin_{\textbf{C}} \sum_{i=1}^k d(\textbf{C}, \textbf{P}_i), 
    \label{eq:objective1}
\end{equation}
In the literature, several measures of $d(\textbf{P}_i, \textbf{P}_j)$ have been discussed including the Rand index, Adjusted Rand index, Fawlkes and Mallows measure, Mirkin distance, variation of information, split-join measure, and the Jaccard index ~\cite{meila2005comparing}. 
In our algorithm, we aim to minimize the Mirkin distance that measures disagreements between two partitions. 
The Mirkin distance between two partitions $\textbf{P}_i$ and $\textbf{P}_j$ counts (twice) the number of vertex pairs $(u,v)$ that are grouped together in one partition but separated into different clusters in the other partition.

{\bf An alternative formulation of the optimization problem.} 
Let $\sigma_v(\textbf{c})$ be an indicator variable that denotes the cluster 
$\textbf{c}$ to which vertex $v$ belongs. 
\[
\sigma_v(\textbf{c}) = 
\begin{cases} 
        1 & \text{iff $v$ is in cluster $\textbf{c}$} \\
        0 & \text{otherwise}
    \end{cases}
\]
Next, we define $\delta_{uv}(\textbf{P})$ to denote the co-clustering distance between two vertices $u$ and $v$ in partition $\textbf{P}$ as follows.
\begin{equation}
\delta_{uv} (\textbf{P}) = \prod_{\textbf{c} \in \textbf{P}} \left[ 1 - \sigma_u(\textbf{c}) \sigma_v(\textbf{c}) \right] = \begin{cases} 
        0 & \text{iff $u$ and $v$ are co-clustered in partition $\textbf{P}$} \\
        1 & \text{otherwise}
    \end{cases}
    \label{eqn:rxyi}
\end{equation}
Here, we defined $\delta_{uv}(\textbf{P})$ as a co-clustering distance instead of a similarity measure used by Filkov and Skienna\cite{filkov2004integrating}. 
Then, the total co-clustering distance $\delta_{uv}$ between $u$ and $v$ is
\begin{equation}
    \delta_{uv} = \sum_{i = 1}^{k} \delta_{uv}(\bf{P_k}).
\end{equation}

\begin{figure}[!t]
    \centering
    \includegraphics[width=0.95\linewidth]{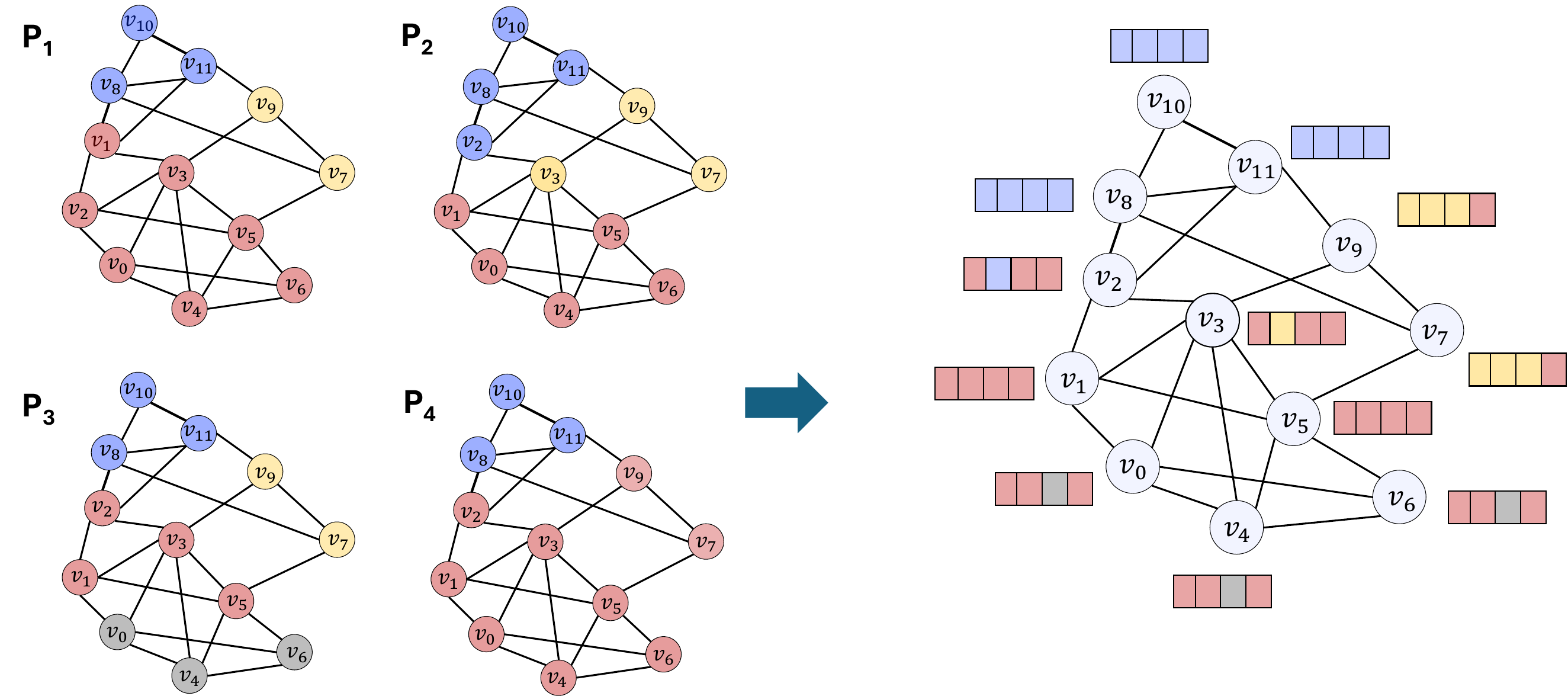}
    \caption{ 
    Four partitions $\textbf{P}_1, \textbf{P}_2, \textbf{P}_3, $ and $\textbf{P}_4$ of a graph with 12 vertices. 
    Within each partition, vertices belonging to different clusters are shown in different colors. 
    The cluster membership vector of each vertex is illustrated in the right figure. 
    Different colors indicate the vertex's membership across the various clusters of the four input partitions.
    The distance between any pair of vertices is the Hamming distance between their vectors. For example, $\delta_{v_0v_1}=1$, $\delta_{v_0v_4}=0$, and $\delta_{v_3v_9}=2$.
    }
    \label{fig:input}
\end{figure}

According to this definition, $\delta_{uv}$ counts the number of partitions in which $(u,v)$ has not been clustered together (which is the Hamming distance between their co-clustering vectors).
Fig.~\ref{fig:input} shows an example graph along with co-clustering distances for different vertices.
We call it the inter-cluster disagreements.
As the Mirkin distance counts the total number of disagreeing pairs of vertices, for this distance metric, minimizing Eq.\ref{eq:objective1} is equivalent to minimizing intra-cluster disagreements and maximizing inter-cluster disagreements.
Hence we represent our 
optimization problem
of Eq.~\ref{eq:objective1} as follows:
\begin{equation}
    \textbf{C}^* = 
    \argmin_{\textbf{C}} 
    \sum_{uv} \left[k - 2 \delta_{uv}\right]  \, \delta_{uv}(\textbf{C})\; .
    \label{eq:objective2}
\end{equation}
The complete derivation of Eq.~\ref{eq:objective2} from Eq.~\ref{eq:objective1} is provided in the appendix.

\subsection{Naive optimization and limitations}
Solving Eq.~\ref{eq:objective2} exactly poses an NP-complete problem for $k>2$ due to the exponential number of potential consensus partitions~\cite{filkov2004integrating}.
Filkov and Skiena~\cite{filkov2004integrating} developed a greedy algorithm to find a consensus partition from a collection of partitions of points in Euclidean space.
Their algorithm pre-computed cluster agreements between every pair of points and stores them in memory to be used in an iterative greedy algorithm.
Similarly, the algorithm by Lancichinetti and Fortunato~\cite{lancichinetti2012consensus} computed agreements between every pair of vertices and used them in a subsequent community detection algorithm. 
However, these methods exhibit three significant limitations in the context of graph consensus clustering.
{\bf (a) High time and memory requirements:} Computing $\delta_{uv}$ for all vertex pairs requires $O(n^2)$ time and memory, where $n$ is the number of vertices. This is too costly and often impractical for large-scale graphs. Additionally, no existing work provides parallel implementations capable of leveraging multi-core processors for quicker consensus generation.
{\bf (b) Overlooking graph structure:} Many existing greedy algorithms for graph consensus clustering are structure-agnostic, as they search for a consensus by considering all pairs of vertices. 
This approach can lead to unnecessary exchanges of vertices among clusters, delaying convergence and compromising the quality of the solution.
{\bf (c) Mixing heterogeneous partitions:} When generating different partitions of a single graph, it is often observed that many of them are very different\cite{calatayud2019exploring, peixoto2021revealing ,kirkley2022representative}. We call the partitions that are similar to each other to be homogeneous and the partitions that are very different from each other to be heterogeneous. Most prior methods constructed a single consensus from all input partitions, potentially mixing heterogeneous partitions - which can undermine the quality of the consensus partition.
Therefore, it is crucial to identify groups of homogeneous partitions and then create a consensus for each homogeneous group.

In this paper, we aim to overcome these limitations by developing a new graph consensus clustering algorithm that rapidly produces a clustering close to the median of all input partitions. 


\begin{figure}[!t]
    \centering
    \includegraphics[width=0.98\linewidth]{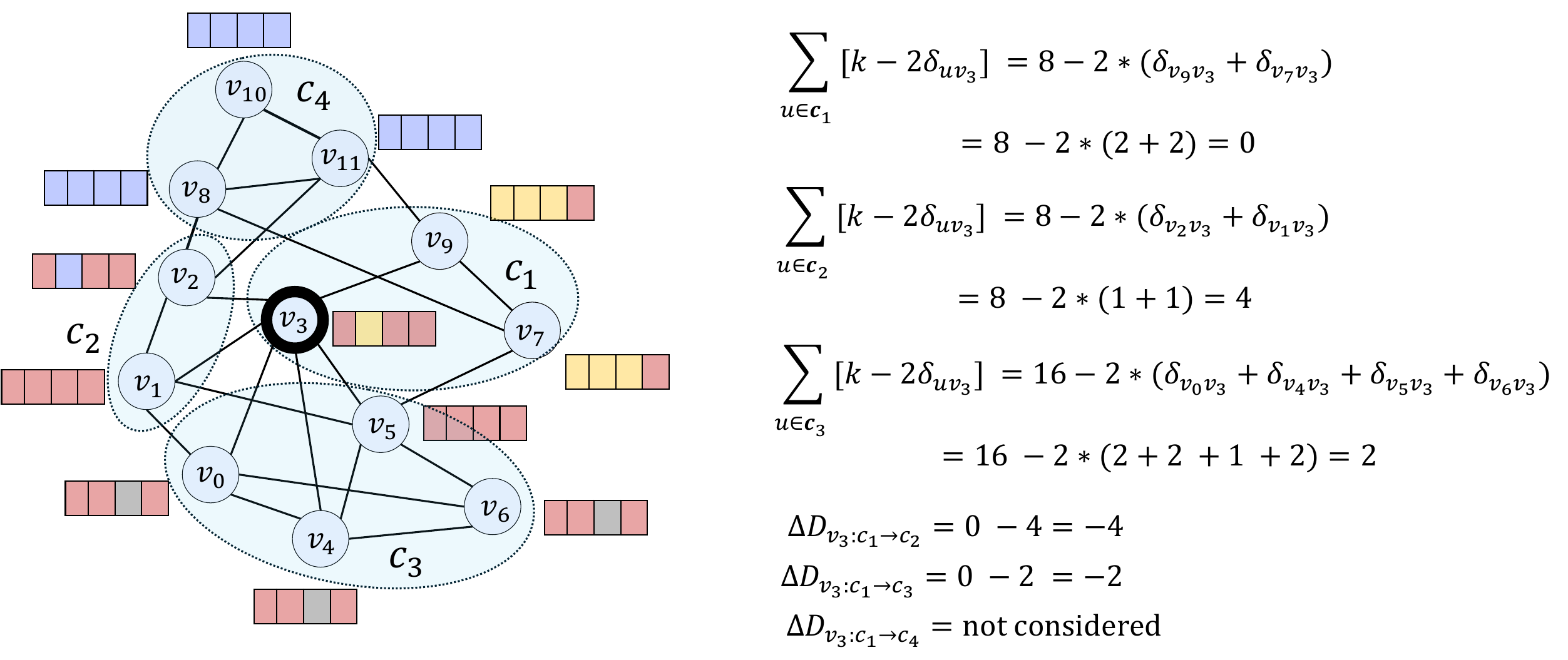}
    \caption{ 
    A consensus clustering $\textbf{C}=\{c_1, c_2, c_3, c_4\}$ of the graph shown in Fig.~\ref{fig:input}. We show the computation to identify the best move for $v_3$ using Eq.~\ref{eq:objective3}. Since $c_4$ does not contain a neighbor of $v_3$, it is not considered in this computation. Based on this calculations, $v_3$ moves to the cluster  $c_2$  since it reduces the total distance by 4. }
    \label{fig:moves}
\end{figure}
\subsection{Graph-aware consensus clustering}
\label{sec:gac}
Given any consensus partition $\textbf{C}$, we design a greedy algorithm that moves vertices across partitions to minimize the overall disagreement among clusters.
The algorithm runs in several iterations.
In each iteration, a vertex is moved from its current cluster to another cluster only if this movement reduces the overall disagreement.
To define this precisely, let the vertex $v$ be in cluster $\textbf{c}$ in the current consensus, with $\textbf{c}'$ being the cluster it might move to.
Then, the change in the objective function is:
\begin{equation}
    \Delta D_{v : \textbf{c} \rightarrow \textbf{c}'}  = \sum_{u} (k - 2 \delta_{uv})\sigma_u(\textbf{c}) - (k - 2 \delta_{uv})\sigma_u(\textbf{c}')  = \sum_{u} [\sigma_u(\textbf{c}')- \sigma_u(\textbf{c})] (2 \delta_{uv} - k ).
     \label{eq:objective3}
\end{equation}
When $\Delta D_{v : \textbf{c} \rightarrow \textbf{c}'}$ is negative, moving $v$ to the new cluster $\textbf{c}'$ reduces the overall disagreements. Hence, $v$ is moved from its current cluster $\textbf{c}$ to another cluster $\textbf{c}'$ only when $\Delta D_{v : \textbf{c} \rightarrow \textbf{c}'}<0$.
Our algorithm does not evaluate all possible clusters 
$\textbf{c}'$ to which the vertex $v$ could move. Instead, we only consider those clusters 
$\textbf{c}'$ where $v$ has at least one neighbor in the graph.

Thus, the cluster $\textbf{c}^*(v)$ where the vertex $v$ might move to is computed as follows: 
\begin{equation}
    \textbf{c}^*(v) = \argmin_{\textbf{c}'\in \textbf{C}_v}  ~ \Delta D_{v : \textbf{c} \rightarrow \textbf{c}'},
    \label{eq:objective4}
\end{equation}
where $\textbf{C}_v$ is the set of all clusters in the current consensus where vertex $v$ has at least one neighbor.






Fig.~\ref{fig:moves} shows the calculations for moving a vertex from its current cluster to other clusters.
Notice that Eq.~\ref{eq:objective4} only considers moving vertices along the edges of the graph, similar to the approach employed in greedy community detection algorithms like the Louvain algorithm~\cite{blondel2008fast}.
Considering the graph structure, at most $O(m)$ moves are considered as opposed to $O(n^2)$ moves in each iteration.
Methods oblivious to the graph structure often try to improve this computational complexity but they still have high space complexity ($O(n^2)$).
Our method does not have such a high memory requirement as well.
We discuss the computation and memory complexity of our algorithm in more detail in Sec.~\ref{subsec:algo}.

{\bf Grouping homogeneous partitions.} We take a simple pre-processing step to group the homogeneous partitions.
At first, we create a complete graph, where each input partition is treated as a vertex, and the distance between them serves as the edge weight.
We used split-join distance~\cite{dongen2000performance} as edge weights for its intuitiveness, but other distance metrics can be used as well.
Next, we sparsify this complete graph by removing edges whose weights are greater than a threshold $\lambda$.
This edge removal step creates multiple connected components where each component is composed of relatively homogeneous partitions with inter-partition distances less than $\lambda$.
We empirically select $\lambda$ using the elbow method, a heuristic for determining the optimal number of clusters in a dataset.
Once homogeneous groups of partitions are identified, we apply the consensus clustering algorithm independently to each group.
This pre-processing step is equivalent to clustering of clustering~\cite{kirkley2022representative} or meta clustering~\cite{caruana2006meta}. We stress that our emphasis here is on the consensus clustering technique, not on the procedure to identify homogeneous sets of partitions. In fact, 
any method that involves clustering of clustering, as discussed in previous studies~\cite{calatayud2019exploring,peixoto2021revealing,kirkley2022representative} can be employed before utilizing our consensus clustering approach. 

\subsection{The algorithm}
\label{subsec:algo}
Given the optimization objectives mentioned in the previous sections, we describe the steps of our algorithm below.

\begin{itemize}
    \item {\bf Input:} A graph $G(V,E)$ with $n$ vertices and $m$ edges. A set of $k$ partitions of the graph $\textbf{P}_1, \textbf{P}_2, ..., \textbf{P}_k$. 
    \item {\bf Find homogeneous groups:} Find a homogeneous group by using the method discussed in Section~\ref{sec:gac}. 
    \item {\bf Initialization:} 
We begin with a consensus partition comprising $n$ clusters, where each vertex constitutes a separate cluster: $\textbf{C} = \{\{v_1\}, \{v_2\}, ..., \{v_n\}\}$.

    \item {\bf Greedy moves:} 
Find the best move for each vertex $v$ in the graph using Eq.~\ref{eq:objective4}. 
$v$ is moved from its current cluster $\textbf{c}$ to $\textbf{c}'$ only if 
$\Delta D_{v : \textbf{c} \rightarrow \textbf{c}'}<0$.

\item {\bf Stopping condition:} 
The algorithm stops when no vertex can be moved.
\end{itemize}

{\bf Computation and memory complexity.}
To identify the homogeneous groups of partitions, we create a complete with $k$ vertices.
This involves comparing $k(k-1)/2$ pairs of partitions, where calculating the split-joint distance between two partitions of $n$ elements requires $O(n \log n)$ time \cite{csardi2006igraph}.
After pruning for a certain value of $\lambda$, this graph contains at most $O(k^2)$ edges, hence detecting connected components on such a graph using any graph traversal technique takes $O(k+k^2)$ time.
Overall, finding homogeneous groups of partition takes $O(k^2n \log n)$ time for a certain value of $\lambda$.

After detecting the homogeneous groups, we make a consensus of each group.
The computational complexity of each iteration of the consensus algorithm depends on the cluster sizes and the number of inter-cluster edges of the consensus partition before the iteration.
Let $b$ represent the number of clusters in the current consensus partition. 
On average, each cluster then contains $\frac{n}{b}$ vertices.
To compute the feasibility of each potential move for each vertex, the algorithm computes two terms on the right-hand side of Eq.~\ref{eq:objective3}.
If there are $k_{h}$ input partitions in the homogeneous group, the computation required for each term is $O(\frac{nk_{h}}{b})$.
If there are $\hat{m}$ inter-cluster edges, the number of potential moves is at most $2\hat{m}$.
Hence, the total computation required on each iteration is $O(\frac{\hat{m}nk_h}{b})$.
One can argue that this complexity can be as high as $O(n^2)$ if $\hat{m}$ is very high and $b$ is very low.
For the simple case of same-size clusters, it is easy to see that both of those conditions cannot occur simultaneously.
As we start from singleton clusters, during the initial iterations $\hat{m}$ would be high but $b$ would be high as well.
As one of the implicit goals of graph clustering is to reduce $\hat{m}$, it decreases in later iterations but $b$ decreases too as small clusters merge with bigger clusters.
In practical cases, i.e. graphs with heterogeneous degree and cluster size distributions, this analysis is not so simple.
Large clusters having many inter-cluster edges may enhance the computational cost.
It might be possible to solve this problem without penalizing the quality too much by appropriate sampling and graph sketching techniques.
We leave that for future work.
However, we present an empirical evaluation of the runtime in Sec.~\ref{subsec:runtime-experiments}.

The median clustering algorithm requires no additional memory beyond what is needed to store the input partitions and the final output.
We have two inputs to our algorithm - the graphs and the input partitions.
Storing the graph takes $O(m)$ space if we store the edges.
Storing the input partitions takes $O(nk)$ space if represented as a $n \times k$ matrix, where each column represents a partition and each row represents cluster memberships of a vertex $v$ across different input partitions(as discussed in Sec.~\ref{subsec:problem}).
Hence the memory complexity is $O(m+nk)$.
It is possible to reduce time complexity by increasing the space complexity, as discussed in Sec.~\ref{sec:gac}.
For this work, we choose to make the algorithm applicable to large realistic graphs through parallel computation keeping the memory requirement the same.
We discuss the parallel algorithm in  Sec.~\ref{sec:parallel} below.

\subsection{Parallel algorithm}
\label{sec:parallel}
The algorithm discussed in the previous section identifies the best possible move for every vertex considering the consensus partition at the beginning of the iteration.
After some validation checks of the identified moves, all moves are performed together at the end of the iteration.
Because one move may change the partition in such a way that some other move may not be optimal anymore, one can argue to identify one best move in an iteration, perform it, and then move to the next iteration.
Some other methods, i.e. Filkov and Skienna\cite{filkov2004integrating}, work in this way.
However, empirically we observe no quality degradation in our experiment by adopting the approach mentioned in the previous section.
In this way, the algorithm takes a few more iterations to converge but they converge to the same quality solution.
Crucially, this approach eliminates sequential dependencies, enabling effortless parallelization of the most computationally demanding components.
As discussed in the previous paragraph, computation of the two terms of Eqn.~\ref{eq:objective3} is the most compute-intensive part.
Because this computation for one potential move does not depend on any other move within an iteration, we can compute for all potential moves in parallel.
We implement our algorithm in C++ and use OpenMP programming model to implement the parallel algorithm.
We assign the computation of every potential move to an OpenMP thread.
We assess the amount of required computation for each move from the cluster sizes and assign it in such a way that the workload is balanced between parallel threads.

%% file: experiments.tex
\section{Results}
We evaluate the performance of our algorithm using three key metrics, which demonstrate its effectiveness both in scenarios with and without community structures and its applicability to large-scale networks:
\begin{enumerate}
    \item {\bf Accuracy}: We demonstrate the accuracy of our algorithm by testing it against ground truth communities in both synthetic and real-world networks.
    \item {\bf Median partition}: We measure how effectively our algorithm identifies a median partition.
    \item {\bf Scalability}: We discuss the capability of our algorithm to handle increasingly large input graphs.
\end{enumerate}

\subsection{Experiment Settings}

\subsubsection{Graphs}
We use synthetic and real-world graphs to evaluate our consensus clustering method empirically.

\textbf{Synthetic graphs}.
We use LFR benchmark graphs of different clustering structures~\cite{lancichinetti2008benchmark}.
To pick the parameters of LFR graphs, we take ideas from the original paper and other works that use this benchmark graph.
In all cases, we fix the following parameters - minimum degree = 5, maximum degree = 50, minimum cluster size = 5, maximum cluster size = 200, degree exponent $\gamma=3.0$, and community size exponent $\beta=1.1$. 
To generate graphs of different pronounced clustering structures, we vary the mixing parameter $\mu$ from 0.1 to 0.7 keeping other parameters fixed.
A smaller value of $\mu$ represents a more pronounced clustering structure in the graph.
To generate graphs of different sizes, we vary the number of vertices from 200 to 125000.
We generate these graphs using the Networkx python package\cite{hagberg2008exploring}.

\textbf{Real-world graphs}.
We use three real-world graphs from two different application scenarios.
\emph{email-Eu-core} is an email communication network containing 1k vertices and 26k edges.
\emph{Levine13} and \emph{Samusik} comes from single cell experiments.
\emph{Levine13} contains 81k vertices and 1.8 million edges while \emph{Samusik} contains 514k vertices and 12 million edges.

\subsubsection{Input partitions}
To generate different partitions of the same graph, we design the experiments in two ways: (1) when partitions are slightly different from each other, and (2) when partitions are significantly different from one another.

\textbf{Slightly different partitions}.
To generate a set of partitions that are not very different from each other, we use 16 different random initialization parameters of the \emph{louvain} algorithm\cite{blondel2008fast} to get 16 different partitions for the same graph.
To generate these partitions, we use the implementation available on the CDLib python package\cite{rossetti2019cdlib}.
We choose 16 different randomization parameters equivalent to 16 different vertex orderings for the updates of the algorithm.
Narrow black error bars of Fig.~\ref{fig:lfr-louvain_n5000_quality} show that these partitions are slightly different.


\textbf{Widely different partitions}.
To generate a set of partitions that may be very different from each other, we use different parameters of 14 different clustering algorithms to get 38 different partitions of the same graph.
We use the following 14 clustering algorithms available in the CDLib python package\cite{rossetti2019cdlib} - 
\emph{label-propagation}\cite{cordasco2010community},  \emph{leiden}\cite{traag2019louvain}, \emph{significance-communities}\cite{traag2013significant}, \emph{surprise-communities}\cite{traag2015detecting}
, \emph{cnm-greedy-modularity}
, \emph{paris}\cite{bonald2018hierarchical}
, \emph{louvain}\cite{blondel2008fast}
, \emph{infomap}\cite{rosvall2008maps}
, \emph{walktrap}\cite{pons2005computing}
, \emph{markov-clustering}\cite{enright2002efficient}
, \emph{expectation-maximization}\cite{newman2007mixture}
, \emph{ricci-community}\cite{ni2019community}
, \emph{stochastic-block-model}\cite{peixoto2014efficient}, 
and \emph{spinglass}\cite{reichardt2006statistical}.
We use the default parameter of most of these implementations other than \emph{louvain}, \emph{markov-clustering}, \emph{expectation-maximization}, \emph{spinglass}, and \emph{ricci-community} algorithms as it was not allowed by the implementation.
We use different resolution and vertex ordering parameters for \emph{louvain}, different pruning and inflation parameters for \emph{markov-clustering}, randomly chosen expected number of clusters for \emph{expectation-maximization} as well as \emph{spinglass} and different probability distribution parameters for \emph{ricci-community}.
The wide black error bars of Fig.~\ref{fig:lfr-all_n5000_quality} indicate that these partitions differ significantly from one another.

\subsubsection{Grouping homogeneous partitions}
\begin{figure}[t!]
    \centering
    \includegraphics[width=.48\linewidth]{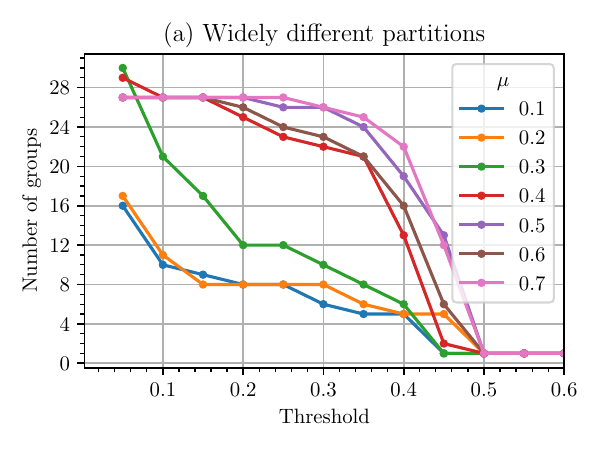}
    \includegraphics[width=.48\linewidth]{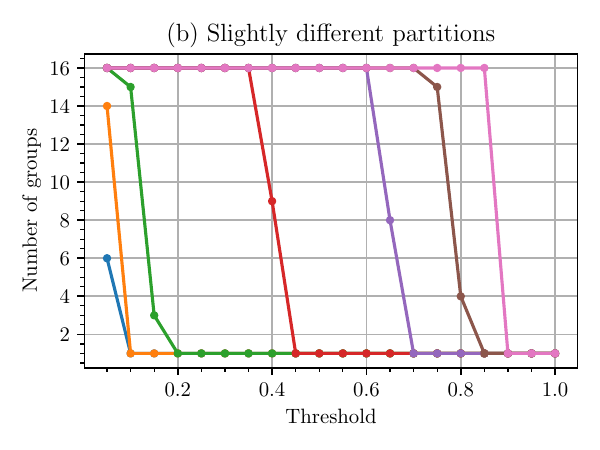}
    \caption{Effect of outlier removal parameter when creating a consensus of (a) widely different partitions and (b) slightly different partitions.
    LFR benchmark networks with n=5000 are used for this experiment.
    For (a), 38 different partitions were involved; for (b), 16 different partitions were involved.}
    \label{fig:outlier-removal-parameter}
\end{figure}
In Sec.~\ref{sec:gac}, we discussed our method of grouping homogeneous sets of partitions.
Fig.~\ref{fig:outlier-removal-parameter} shows the effect of the distance threshold parameter $\lambda$.
We change the distance threshold value from 1.0 down to 0.05 and observe how many groups are detected and the size of the groups.
We observe that as the distance threshold increases, the number of groups decreases.
For our experimental dataset, we also observe that while most of these groups contain very few partitions, one group usually is significantly larger than the others.
Based on this study, we pick the distance threshold parameter ($\lambda$) for subsequent experiments.
Since our experimental dataset primarily consists of one large group and several small groups, we select the value of $\lambda$
at the point where more than one group begins to emerge (e.g., $\lambda=0.5$ in Fig.~\ref{fig:outlier-removal-parameter}(a)). 
We then take the partitions in the largest group and determine their consensus.



\subsubsection{Baseline algorithms}
We compare our algorithm, \emph{median-consensus}, with the following baseline algorithms.

\textbf{\emph{lf-consensus}}\cite{lancichinetti2012consensus}.
We use \emph{lf-consensus} as the baseline algorithm to perform consensus clustering on graphs.
In each iteration, the original \emph{lf-consensus} work runs the same algorithm multiple times on the consensus graph generated from the previous iterations.
To be consistent with that, in each iteration, we use the same parameters of the same algorithms that have been used to generate the input partitions.
We observed that for experiments involving widely-different partitions, \emph{lf-consensus} does not converge.
Hence, we omit this method in such cases.

\textbf{\emph{kirkley-newman}}\cite{kirkley2022representative}.
We use \emph{kirkley-newman} algorithm to pick the best candidate partition from the inputs as a consensus.
\emph{kirkley-newman} algorithm groups the partitions into homogeneous groups and picks one representative partition from each group.
To be consistent with our experiments where we make consensus of the largest homogeneous group of partition, we pick the representative of the largest group found by \emph{kirkley-newman} as the consensus partition.

\textbf{\emph{boem}}\cite{filkov2004integrating}.
As our method focuses on minimizing the distance from the consensus partition to all input partitions, we compare it with other methods optimizing the similar goal.
We use \emph{boem} proposed by Filkov and Skiena as a baseline algorithm from this category.
Since \emph{boem} is not designed for graphs, we apply it without utilizing the graph structure.

\subsubsection{Evaluation metrics}
We evaluate partitions with several different partition comparison metrics.
Gates et al. compiled a summary of the performance of the most popular partition comparison metrics, categorized by their desirable properties~\cite{gates2019element}.
Meila proved that it is impossible to have any partition distance metric that satisfies all of the most desirable properties of partition comparison~\cite{meila2005comparing}.
Therefore, we select three popular partition comparison metrics, each of which satisfies a subset of the most desirable properties - (i) \emph{variation of information distance}\cite{meila2007comparing} that quantifies the amount of information lost or gained when switching from one partition to another, (ii) \emph{split-join distance}\cite{dongen2000performance} that quantifies the minimum number of split and join operations required to transform one partition into the other and (iii) \emph{rand distance}\cite{rand1971objective} that quantifies the number of pairs of elements that are clustered differently in the two partitions.
Please note that \emph{rand distance} is a normalized form of \emph{mirkin distance}, hence they are equivalent. However, \emph{rand distance} can be perceived as probability and hence more intuitive.
All three of these measures are true distance metrics and a combination of these measures covers most of the desirable properties of cluster comparison.
Our motivation is that observing good performance in terms of all three of these metrics would strengthen the argument.

\subsection{Accuracy}
\label{subsec:gt-comparison}
When the ground-truth communities are known, we evaluate the results by measuring how closely the consensus partition matches the ground truth.
We evaluate this by computing the distance from the ground-truth partition to the other partitions.
The partition that is closest to the ground truth is considered to be best in terms of accuracy.
The ground-truth partitions of the LFR benchmark graphs are known by design.
Two of our real-world datasets also have ground-truth information. 

\begin{figure*}[t!]
    \centering
    \includegraphics[width=1.0\textwidth]{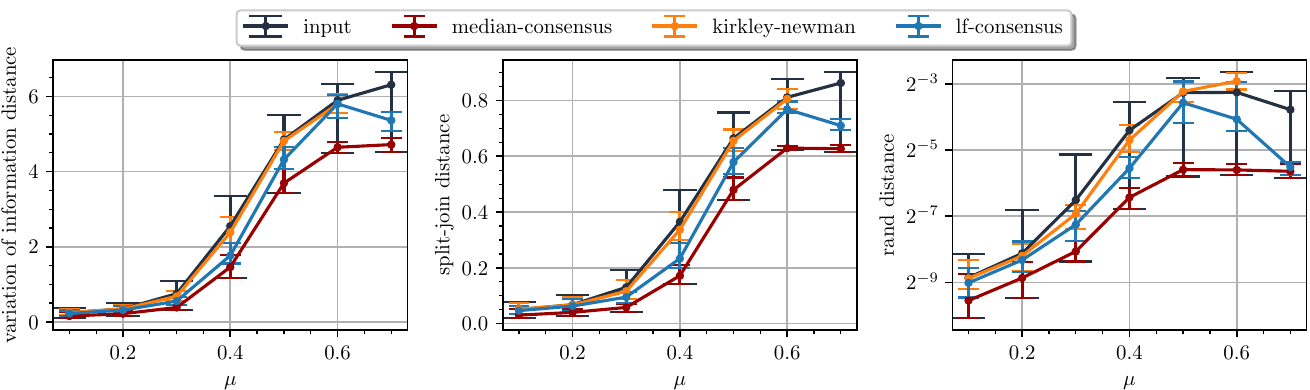}
    \caption{ 
    Accuracy of different consensus partitions for LFR benchmark graphs with n=5000, when slightly different input partitions are considered.
    Here input partitions are obtained by different runs of the Louvain algorithm.
    Since the ground truth communities are known for LFR benchmark graphs, partition accuracy is measured by the distance from the ground truth community (where lower distance values indicate better accuracy).
    The black line represents the accuracy of the input partitions used to build the consensus, while the colored lines represent various consensus methods.
    Points on each line represent the average from 10 sets of experiments, with the corresponding error bars indicating the maximum and minimum values observed across all 10 sets.
    }
    \label{fig:lfr-louvain_n5000_quality}
\end{figure*}

\textbf{Slightly different partitions of synthetic graphs}.
We present our evaluation in Fig.~\ref{fig:lfr-louvain_n5000_quality}.
Across all evaluation metrics, our \emph{median-consensus} consistently identifies a more accurate partition compared to all baseline consensus methods.
Additionally, we observe that in this setting, the consensus produced by \emph{kirkley-newman} closely follows the input partitions. 
We attribute this to the fact that the input partitions are only slightly different from one another, and since \emph{kirkley-newman} selects a solution directly from the input, it tends to follow the overall trend of the input partitions.

\begin{figure*}[t!]
    \centering
    \includegraphics[width=1.0\textwidth]{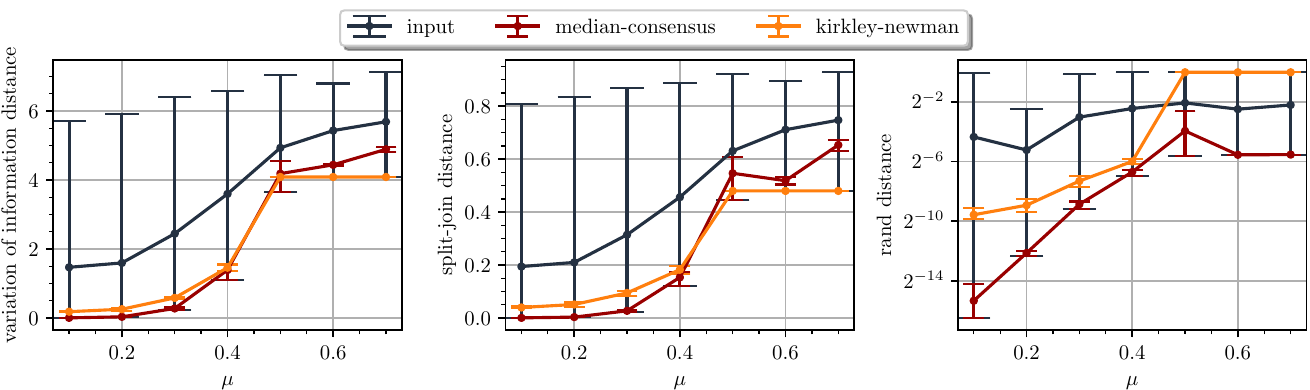}
    \vspace{-10pt}
    \caption{
    Accuracy of different partitions of LFR benchmark graphs with n=5000 when widely different partitions are considered.
    Here input partitions are obtained by different parameters of 14 different clustering algorithms.
    The black line represents the accuracy of the input partitions used to build the consensus, while the colored lines represent different consensus methods.
    Points on each line represent the average from 10 sets of experiments, with the corresponding error bars indicating the maximum and minimum values observed across all 10 sets. 
    In all cases, a lower value in the Y-axis is better as it signifies the distance to ground truth is smaller.
    }
    \label{fig:lfr-all_n5000_quality}
\end{figure*}


\textbf{Widely different partitions of synthetic graphs}.
We present our evaluation in Fig.~\ref{fig:lfr-all_n5000_quality}.
\emph{lf-consensus} did not converge for this case, so we have excluded it from our analysis.
We observe that when the graph exhibits a clear clustering structure
($\mu < 0.5$), across all evaluation metrics, \emph{median-consensus} 
consistently generates more accurate partitions than other baseline methods across all evaluation metrics.
When the graph lacks a significant clustering structure ($\mu > 0.5$), 
it is difficult to determine which method produced more accurate partitions, as the three evaluation metrics yielded conflicting results. 
However, since this evaluation is based on synthetic graphs and real-world graphs with appreciable community structure, we disregard the cases for $\mu > 0.5$.
As the input partitions in this setting differ widely, \emph{kirkley-newman} chooses from the best partitions of the input.
Hence, unlike the case of slightly different partitions, it does not stay close to the average input partition trend.
By comparing Fig.~\ref{fig:lfr-louvain_n5000_quality} and Fig.~\ref{fig:lfr-all_n5000_quality}, we observe that \emph{median-consensus}, which aggregates partitions from various clustering algorithms, yields more accurate partitions across all three evaluation metrics than the consensus of partitions produced by Louvain alone.
Hence, we argue that aggregating partitions from multiple algorithms could be advantageous following outlier removal.

\begin{figure*}[t!]
    \centering
    \begin{subfigure}[t]{0.48\textwidth}
        \includegraphics[width=0.30\linewidth]{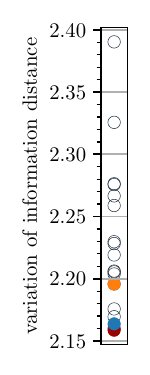}
        \includegraphics[width=0.30\linewidth]{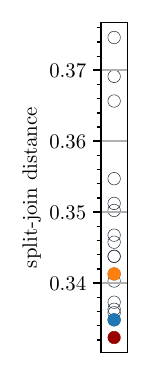}
        \includegraphics[width=0.30\linewidth]{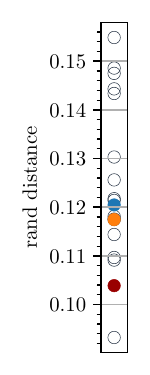}
        \caption{
            Accuracy of clustering the \emph{email-Eu-core}\cite{leskovec2007graph} communication graph.
            Grey circles represent the clusters found by different runs of the Louvain algorithm.
            Red, blue, and orange dots represent the clustering found by \emph{median-consensus}, \emph{lf-consensus} and \emph{kirkley-newman}  respectively.
            In all cases, lower is better.
        }
        \label{fig:email-eu-core}
    \end{subfigure}
    ~
    \begin{subfigure}[t]{0.48\textwidth}
        \includegraphics[width=0.30\linewidth]{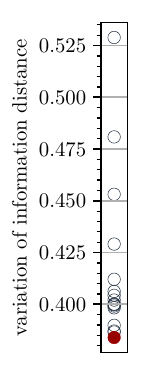}
        \includegraphics[width=0.30\linewidth]{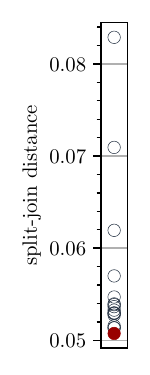}
        \includegraphics[width=0.30\linewidth]{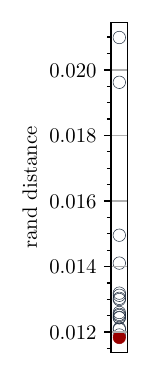}
        \caption{
            Accuracy of clustering kNN graph of \emph{Levine13}\cite{weber2016comparison} single-cell data from FastPG\cite{bodenheimer2020fastpg} pipeline.
            Grey circles represent the clusters found by different runs of the Louvain algorithm.
            Red dots represent the partition found by \emph{median-consensus}.
            In all cases, lower is better. Other baseline methods failed to generate a consensus for this large graph.
        }
    \label{fig:levine13}
    \end{subfigure}

    \caption{Accuracy of consensus clustering in two real-world networks.
    }
    \label{fig:real-gt}
\end{figure*}
\textbf{Clustering a real-world communication network}.
We first demonstrate the applicability of our consensus clustering algorithm on the \emph{email-Eu-core} graph - which is an email communication network of a large European research institute\cite{leskovec2007graph}.
Each vertex represents an individual belonging to some department of the research institute.
A group of individuals belonging to the same department is considered to be a ground truth cluster in the graph.
We generate different partitions of the network by applying the Louvain algorithm several times with different random vertex orderings and then apply consensus.
We compare the different partitions with the known ground truth information from the metadata.
Fig.~\ref{fig:email-eu-core} shows that \emph{median-consensus} produces a more accurate partition than the other baseline methods.

\textbf{Clustering single-cell experiment data.}
We also demonstrate the applicability of our consensus clustering algorithm in real-world single-cell experiments.
FastPG is a pipeline of clustering millions of cells from single-cell experiments\cite{bodenheimer2020fastpg}.
It has been demonstrated to produce accurate results a magnitude faster than other comparative methods.
The superior performance comes from the construction of a k-nearest-neighbor (kNN) graph and applying a variant of the Louvain algorithm to exploit multi-core parallelism.
Because Louvain and its parallel variants suffer from producing inconsistent clustering output for different runs, we applied our consensus clustering algorithm to get a consistent result.
We take the kNN graph generated by the FastPG pipeline for a dataset coming from human bone marrow cells of healthy donors (\emph{Levine13})\cite{weber2016comparison}.
We run the Louvain algorithm with 16 random vertex ordering to get 16 different partitions.
We make a consensus of these 16 different partitions with our \emph{median-consensus} method.
As this dataset contains ground truth cell types, we consider each cell type as a cluster and compare it with partitions.
Fig.~\ref{fig:levine13} shows that in this way we could recover more accurate clusters from the kNN graph.

\subsection{Median partition}
\begin{figure*}[t!]
    \centering
    \includegraphics[width=1.0\linewidth]{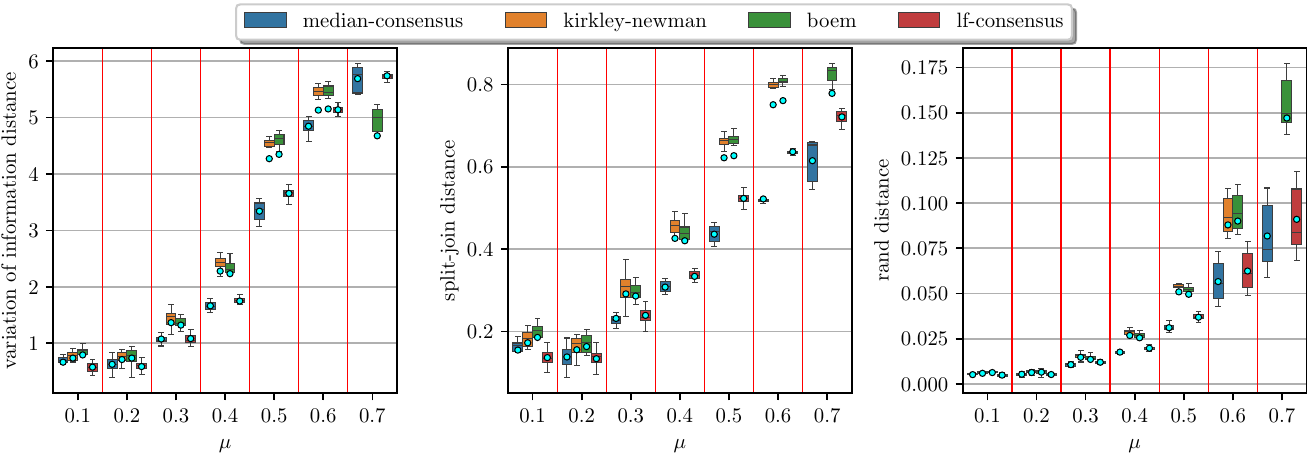}
    \caption{
    Distribution of distance from consensus partition to all input partitions for slightly different partitions.
    LFR benchmark graphs with n=5000 are used.
    Cyan dots represent the mean of each distribution.
    }
    \label{fig:lfr-louvain-n5000-median}
\end{figure*}

\begin{figure*}[t!]
    \centering
    \includegraphics[width=1.0\linewidth]{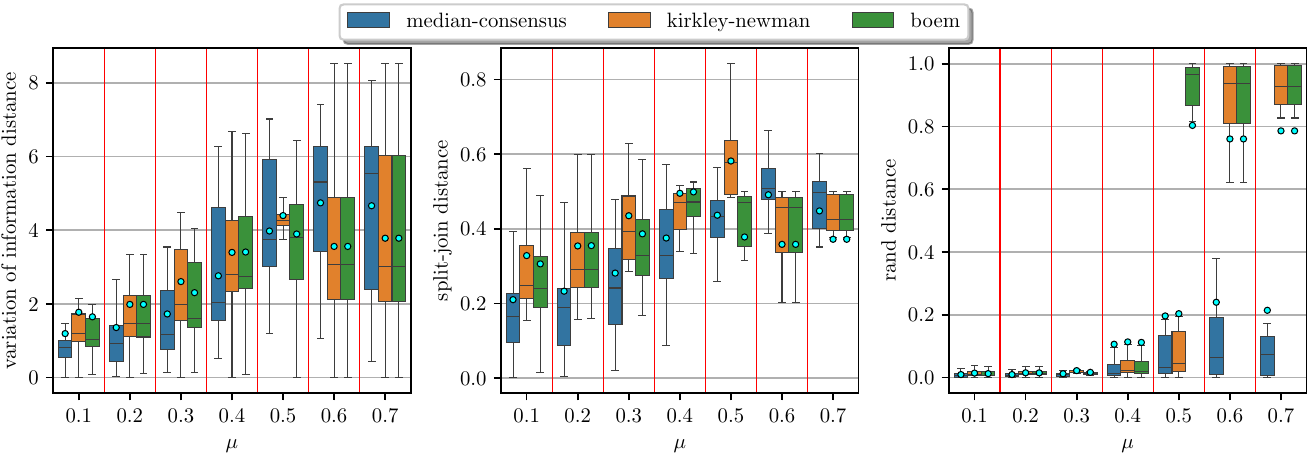}
    \caption{
    Distribution of distance from consensus partition to all input partitions for widely different partitions.
    LFR benchmark graphs with n=5000 are used.
    Cyan dots represent the mean of each distribution.
    }
    \label{fig:lfr-n5000-median}
\end{figure*}
In the absence of known ground-truth clusters, we evaluate the consensus partition from a median perspective, as it provides the most impartial assessment.
In the following experiments, we disregard the ground-truth information of the LFR benchmark graphs and instead evaluate the consensus based on its proximity to all input partitions.
For each consensus partition, we calculate its distance to all input partitions and identify the best consensus partition as the one that minimizes the mean distance to the inputs.
For this evaluation, we include \emph{boem} into our consideration as it actively optimizes the goal of finding a median partition.
\emph{lf-consensus} did not converge for widely varying partitions, so it is excluded from consideration in that scenario.
We present our evaluation in Fig.~\ref{fig:lfr-louvain-n5000-median} and ~\ref{fig:lfr-n5000-median}.
Across all distance metrics, we observe that both \emph{median-consensus} and \emph{lf-consensus} performed well in finding a good median consensus partition.
For the graphs that have a very pronounced clustering structure (very small values of $\mu$), \emph{lf-consensus} performed slightly better than \emph{median-clustering} but the latter method worked well across graphs of different clustering structures as well as widely-different partitions.

\subsection{Runtime}
\label{subsec:runtime-experiments}


\begin{figure*}[t!]
    \centering
    \begin{subfigure}[t]{0.55\textwidth}
        \includegraphics[width=0.98\linewidth]{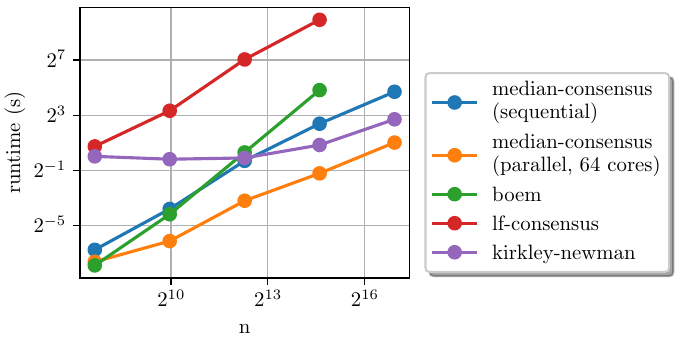}
        \caption{
            Runtime of different consensus methods on LFR benchmark graphs of varying sizes.
            \emph{boem} and \emph{lf-consensus} experiments crashed by going out of memory for n=125000.
        }
        \label{fig:lfr-rt}
    \end{subfigure}
    ~
    \begin{subfigure}[t]{0.35\textwidth}
        \includegraphics[width=0.98\linewidth]{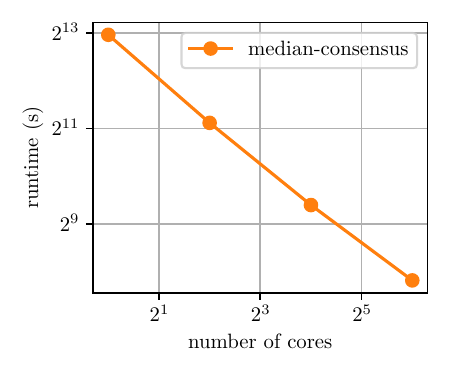}
        \caption{
            Consensus clustering runtime of kNN graph of \emph{Samusik-all}\cite{weber2016comparison} single-cell data from FastPG\cite{bodenheimer2020fastpg} pipeline (514k vertices and 12 million edges) as we increase the number of computing resources.
            Sequential implementation of \emph{kirkley-newman} did not finish in 10 hours.
        }
    \label{fig:samusik-scaling}
    \end{subfigure}

    \caption{
    Runtime and scalability of consensus clustering algorithms.
    }
    \label{fig:rt}
\end{figure*}

We evaluate the runtime of consensus methods in two different ways - (i) runtime on varying-size graphs and (ii) scalability on large practical graphs.
For the first case, we generate LFR benchmark graphs of 200 to 125000 vertices with mixing parameter $\mu=0.4$ and generate different partitions by running the Louvain algorithm multiple times with different vertex ordering.
All our experiments were performed on a computing node equipped with AMD EPYC 7783 processor having 64 cores in a single socket and 256 GB of memory.
In Fig.~\ref{fig:lfr-rt} we show the runtime of each consensus method as the size of the graph increases.
We observe that both \emph{boem} and \emph{lf-consensus} failed to run on the 125000 node graph due to going out of memory.
We also observe that even though \emph{kirkley-newman} ran faster than the sequential version of \emph{median-consensus}, by utilizing high-performance computing resources efficiently the parallel version of \emph{median-consensus} ran several times faster than \emph{kirkley-newman} for larger graphs.
On a larger real-world graph (Fig.~\ref{fig:samusik-scaling}) we observe that \emph{median-consensus} is the better performing algorithm than \emph{kirkley-newman}.
We take a kNN graph of 514k vertices and 12 million edges obtained from single-cell experiments~\cite{bodenheimer2020fastpg} and generate different partitions in the same way as above.
We observe that while \emph{median-consensus} could successfully find consensus in little more than 2 hours using a single computing core, \emph{kirkley-newman} could not finish in 10 hours.
We also observe that the \emph{median-consensus} method shows a linear speedup as we go from 1 computing core to 64 cores.
Utilizing 64 computing cores \emph{median-consensus} produced the same result in less than 4 minutes.
It underscores the fact that \emph{median-consensus} method is capable of producing consensus of even larger datasets utilizing the high-performance computing resources of modern days.

%% file: conclusion.tex
\section{Conclusions}
This paper introduces a new algorithm for finding a consensus among different partitions of a complex network. Unlike previous graph consensus clustering methods, our algorithm aims to find a median of all input partitions. We identified several key advantages of the median partitions generated by our algorithm.
(1) A median partition is typically closer to ground truth communities compared to methods that do not search for median partitions.
(2) Our median partition provides a consensus with fewer disagreements among input partitions.
(3) By using a graph-aware greedy algorithm, we reduce the computational complexity of the process.
(4) We eliminate sequential dependencies among vertex movements and develop a parallel algorithm for multi-core processors, achieving linear speedup as the number of processing cores increases.
(5) Due to faster computations and reduced memory requirements, our median consensus algorithm can handle large graphs where other methods fail to generate any consensus.
(6) We also developed a simple step to group homogeneous partitions before independently finding the consensus of each group.
Overall, our work enables finding consensus partitions on large graphs quickly and accurately.

An extension of our algorithm would be to construct consensus clustering for dynamic networks that evolve over time. 
In such networks, clustering structures change as the network evolves, making consensus clustering an effective approach for identifying core and stable clusters~\cite{hussain2022disruption, yadav2020resilience, dong2018resilience}. 
Since our algorithm identifies a median partition that, on average, minimizes distance with all input partitions, it can be readily adapted to dynamic networks if a suitable distance metric is defined for dynamic graphs. We plan to explore this promising research direction in the future

\section*{Data availability} Our implementation of \emph{median-consensus} is located at \href{https://github.com/taufique71/pamcon}{https://github.com/taufique71/pamcon}.
We used synthetic and publicly available networks for our experiments. The sources of the networks are mentioned in their respective results sections.

\section*{Acknowledgements}
MTH and AA are supported by the Applied Mathematics Program of the DOE Office of Advanced Scientific Computing Research under contracts numbered DE-
SC0022098 and DE-SC0023349 and by the NSF OAC-2339607 grant. MH, SC, and MTH (internship at Pacific Northwest National Laboratory (PNNL)) were supported by the DOD SERDP RC20-1183 grant titled ``Networked Infrastructures under Compound Extremes.'' PNNL is operated by Battelle Memorial Institute for the U.S. Department of Energy under Contract No. DE-AC05-76RL01830.  SF was partially supported by the National Institutes of Health under awards U01 AG072177 and U19 AG074879. FR was partially supported by the Air Force Office of Scientific Research under awards FA9550-21-1-0446 and FA9550-24-1-0039. The funders had no role in study design, data collection, and analysis, the decision to publish, or any opinions,
findings, conclusions, or recommendations expressed in the
manuscript.

\section*{Author contributions} The algorithm was developed through collaborative discussions among all authors. MTH implemented the algorithms and conducted the experiments for the paper. All authors contributed to the writing.
\section*{Competing interests}
The authors declare no competing interests.

%% file: Appendix.tex
\appendix
\section{Appendix}

\begin{figure}[h]
    \centering
    \includegraphics[width=0.5\linewidth]{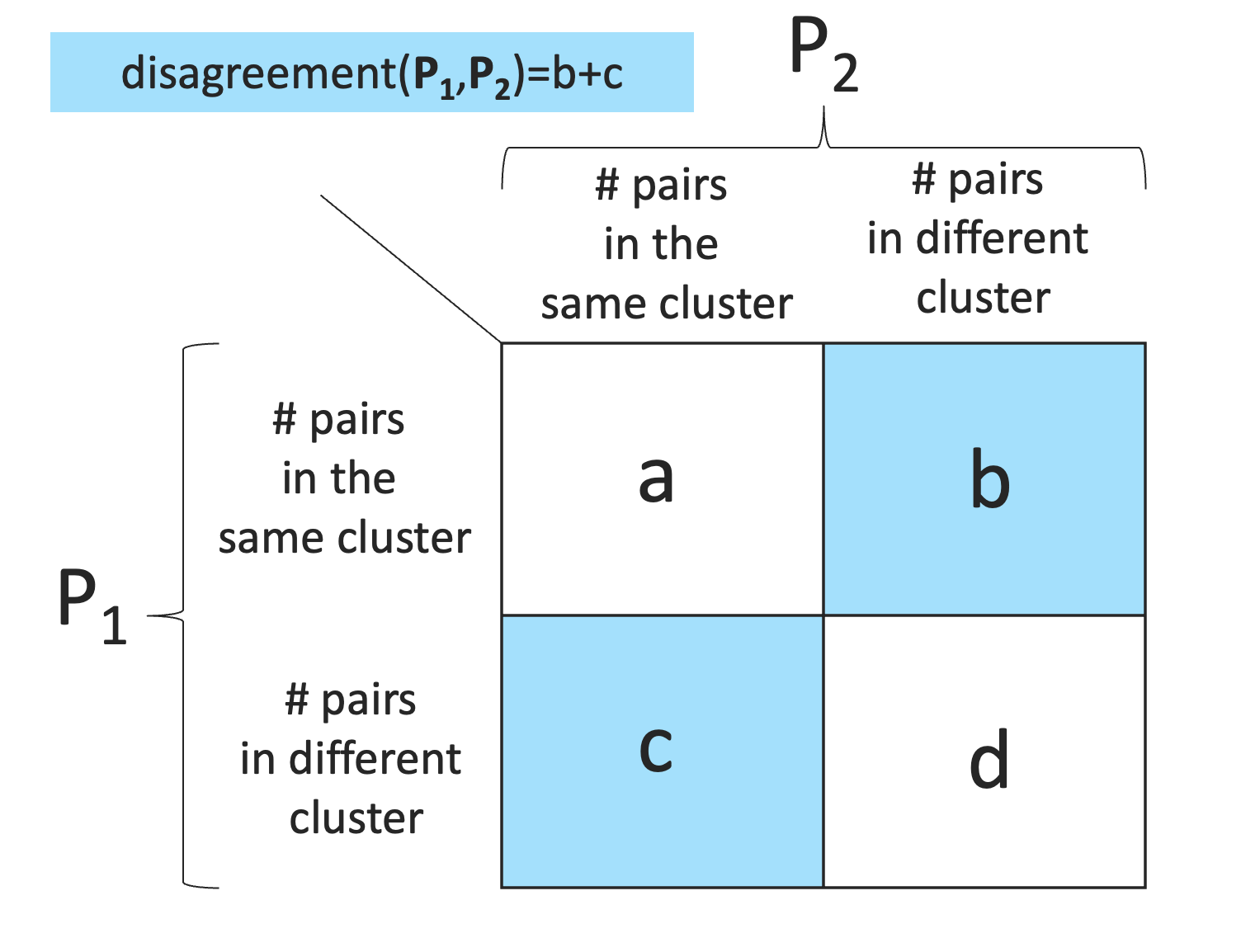}
    \caption{ Contingency table of number of vertex pair count when comparing two clusterings.
    Pairs of vertices for a partion can be divided into two groups - pairs belonging to the same cluster and pairs belonging to different clusters.
    When comparing two clusterings the pairs can be divided into four groups - a, b, c and d.
    }
    \label{fig:disagreement-scheme}
\end{figure}
Equation~\ref{eq:objective1} can be optimized with any distance metric. In our algorithm,
we aim to minimize the Mirkin distance that measures disagreements between two partitions. 
Here, we show the derivation of Equation~\ref{eq:objective2} from Equation~\ref{eq:objective1} using the Mirkin distance.
The Mirkin distance is a measure used in clustering and partitioning contexts to quantify the dissimilarity between two partitions of a set. It is defined as the number of pairs of elements that are clustered together in one partition but not in the other.
Hence, according to Fig.~\ref{fig:disagreement-scheme}, the Mirkin distance $\textbf{P}_1$ and $\textbf{P}_2$ is $d(\textbf{P}_1, \textbf{P}_2) = b+c$.
This distance is a metric and is equivalent to the symmetric difference between the sets of all co-clustered pairs in $\textbf{P}_1$ and $\textbf{P}_2$ (i.e. the symm. diff. of the equivalence relations induced by the partition).
Generally, the Mirkin distance between $\textbf{P}_i$ and $\textbf{P}_j$ is defined as 
\begin{equation}
    d(\textbf{P}_i, \textbf{P}_j) = b+c = \sum_{uv} (\delta_{uv}(\textbf{P}_i) + \delta_{uv}(\textbf{P}_j) - 2\delta_{uv}(\textbf{P}_i) \delta_{uv}(\textbf{P}_j)).
    \label{eqx:distance}
\end{equation}
In this derivation, we used the following properties of a symmetric difference $A \triangle B$ between two sets $A$ and $B$: $|A \triangle B| = |A| + |B| - 2|A \cap B|$.

We now derive Equation~\ref{eq:objective2} from Equation~\ref{eq:objective1}.
\begin{align*}
\sum_{i = 1}^{k} d(\textbf{C}, \textbf{P}_i) &= \sum_{i = 1}^{k} \sum_{uv} \left(\delta_{uv}(\textbf{C}) + \delta_{uv}(\textbf{P}_i) - 2\delta_{uv}(\textbf{C}) \delta_{uv}(\textbf{P}_i) \right) \\
& = \sum_{uv} ( \sum_{i = 1}^{k} \delta_{uv}(\textbf{C}) + \sum_{i = 1}^{k} \delta_{uv}(\textbf{P}_i) - 2 \sum_{i = 1}^{k}  \delta_{uv}(\textbf{C})  \delta_{uv}(\textbf{P}_i)) \\
& = \sum_{uv} (k \delta_{uv}(\textbf{C}) + \delta_{uv} - 2\delta_{uv}(\textbf{C}) \delta_{uv} ) \\
& = \sum_{uv} \delta_{uv} + \sum_{uv} [k - 2 \delta_{uv}]  \delta_{uv}(\textbf{C}) \\
& \approx \sum_{uv} [k - 2 \delta_{uv}]  \delta_{uv}(\textbf{C})
\end{align*}
 Note that $\sum_{uv} \delta_{uv}$ does not depend on $\textbf{C}$, thus we do not consider it in our optimization problem.



Now, we consider the change in the objective function when the vertex $v$ is moved from its current cluster $\textbf{c}$ to another cluster $\textbf{c}'$.

\begin{align*}
  \Delta D_{v : \textbf{c} \rightarrow \textbf{c}'} & = 
 \sum_{u\in \textbf{c}} (k - 2 \delta_{uv})- \sum_{u \in \textbf{c}'} (k - 2 \delta_{uv}) \\
 & = \sum_{u} (k - 2 \delta_{uv})\sigma_u(\textbf{c}) - (k - 2 \delta_{uv})\sigma_u(\textbf{c}') \\
& = \sum_{u} [\sigma_u(\textbf{c}')- \sigma_u(\textbf{c})] (2 \delta_{uv} - k )
\end{align*}